\documentclass[12pt]{article}
\usepackage{graphicx}
\usepackage{float}
\usepackage{amsfonts}
\begin{document}

\title{\bf A Study of Holographic Dark Energy Models in Chern-Simon Modified Gravity}

\author{M. Jamil Amir\thanks{mjamil.dgk@gmail.com}$~~$and Sarfraz Ali\thanks{sarfraz270@yahoo.com}
\\Department of Mathematics, University of Sargodha, Pakistan,\\
Department of Mathematics, University of Education, Lahore,\\Pakistan.\\}

\date{}

\maketitle

\begin{abstract}
This paper is devoted to study some holographic dark energy models in the context
of Chern-Simon modified gravity by considering FRW universe. We analyze the equation of state parameter
using Granda and Oliveros infrared cut-off proposal which describes the accelerated expansion of the universe
under the restrictions on the parameter $\alpha$. It is shown that for the accelerated expansion phase
$ -1<\omega_{\Lambda}<-\frac{1}{3}$, the parameter  $\alpha$ varies  according as $1<\alpha<\frac{3}{2}$.
Furthermore, for $0<\alpha<1$, the holographic energy and pressure density illustrates phantom-like theory of the
evolution when $\omega_{\Lambda}<-1$. Also, we discuss the correspondence between the quintessence, K-essence,
tachyon and dilaton field models and holographic dark energy models on similar fashion.
To discuss the accelerated expansion of the universe, we explore the potential and the dynamics of
quintessence, K-essence, tachyon and dilaton field models.
\end{abstract}

Dark Energy Models.

\section{Introduction}
The physicist and cosmologist are facing two fundamental curious  problems, the "dark energy (DE)" and "dark matter (DM)".
Since last decade, the astronomical observational data collected from large scale structures, type Ia Supernovae
and the cosmic microwave background anisotropy supported that our universe is in accelerated expansion
\cite{[1]}-\cite{[3]}. Type Ia supernovae observational data provided the evidences that our universe is
under accelerated expansion due to an exotic energy which has negative pressure and it is so-called DE.
According to the astrophysical observations \cite{[4]}, more than $95$ percent of the contents of our
universe are consist of DM and DE while only about 4 percent is byronic matter
with negligible amount of radiation. It is more interesting that about 70 percent of the energy
density is DE which is responsible of accelerated expansion of the universe.
Although, a huge number of efforts have been made to resolve these issues but there is no satisfactory answer obtained till now.

A number of DE models have been discussed on the holographic principle available in literature \cite{[5]}, \cite{[6]}.
Gao et al. purposed some cosmological constraints on the holographic Ricci dark energy.
Adabi et al. \cite{[7]} discussed the correspondence between the ghost dark energy model and Chaplygin scalar field in
the framework of general relativity (GR). They investigated FRW universe containing DE and DM.
K. Karami and Fehri \cite{[8]}  found the evolution equation as well as equation of state (EoS) parameters
using holographic dark energy (HDE) model with Granda and Oliveros cut-off.  Jamil et al. \cite{[28]}
studied the HDE problem with a varying gravitational constant, in flat and non-flat universe. Alongwith Setare he \cite{[29]}
discussed the HDE issue with a varying gravitational constant, in H\"{o}rava-Lifshitz gravity. With his collaborators
\cite{[30]}, they investigated the model of interacting DE and derive its EoS and found the correspondence between the K-essence, tachyon and dilaton scalar fields with the interacting entropy corrected new agegraphic DE in the non-flat FRW universe.
Jamil et al.  \cite{[9]}  also,  using Granda-Oliveros cut-off, studied the holographic dark energy model in the framework of Brans-Dicke gravity theory.
Many other DE models have been investigated in different theories, for example, DE modal with quintessence \cite{[10]}, quintom field \cite{[11]}, K-essence field \cite{[12]}, tachyon field \cite{[13]}, dilaton field \cite{[14]}, phantom  field \cite{[15]}.

The cosmic baryon asymmetry is longstanding problem of cosmology which suggests a modification in the theory of
GR by introducing Chern-Simons (CS) term in inflationary process \cite{[16]}.
The CS modified gravity is an extension to GR introduced by Jackiw and Pi \cite{[17]}. In this theory,
the gravitational field is coupled with a scalar field using a parity-violating CS term.
Pasqua et al. \cite{[20]}  investigated the HDE model using Granda-Oliveros cut-off, modified holographic
Ricci dark energy model as well as they investigated a model containing higher derivatives of the Hubble
parameter in the context of CS modified gravity. Jamil and Sarfraz \cite{[21]} found the Ricci dark energy
of Amended FRW universe in the frame work of CS modified gravity.

We organize this paper in following order. The brief review of CS modified gravity is presented in section $2$.
In section $3$, we investigate the HDE model and explored the EoS parameter in the framework of CS modified gravity.
The Correspondence between holographic and scalar field models is studied in section $4$. The results are summarized in the last section.

\section{Brief Review of CS Modified Gravity}
The Einstein-Hilbert action  for  CS modified gravity theory is given by \cite{[17]}
\begin{eqnarray}
S=\int d^{4}x\sqrt{-g}[\kappa R+\frac{\alpha}{4}\Theta~^{*}RR-\frac{\beta}{2}(g^{\mu\nu}\nabla_{\mu}\Theta\nabla_{\nu}\Theta+2V[\Theta])]+S_{mat},
\end{eqnarray}
where $\kappa = \frac{1}{16\pi G}$,  $\nabla_{\mu}$ is the covariant derivative, $R$ is the Ricci scalar, $^{*}RR$
is called Pontryagin term  defined as $ ^{*}RR= {{^{*}R^a}_b}^{cd} {R^b}_{acd}$, is topological invariant.
The ${R^b}_{acd}$ is the Reimann tensor and ${{^{*}R^a}_b}^{cd}$ is the dual Reimann tensor defined as ${{^{*}R^a}_b}^{cd}=\frac{1}{2}\epsilon^{cdef}{R^a}_{bef}$.
The terms $\alpha$ and $\beta$ are defined as coupling constants and the function $\Theta$ is called CS coupling field, a function of spacetime using as a deformation function. If  function $\Theta$ is taken to be a constant, CS modified theory reduces to GR identically.

Now, the variation of the action with respect to  metric tensor $g_{\mu\nu}$ and scalar field $\Theta$
yields two field equations of CS modified gravity \cite{[17]}
\begin{eqnarray}
G_{\mu\nu}+l C_{\mu\nu} &=& \kappa T_{\mu\nu},\\
g^{\mu \nu}\nabla_{\mu}\nabla_{\nu}\Theta &=&-\frac{\alpha}{4}~^{*}RR,
\end{eqnarray}
where $G_{\mu\nu}$ is the Einstein tensor, the term $l$ is 4D coupling constant, $C_{\mu\nu}$ is the C-tensor  defined as
\begin{eqnarray}
C^{\mu\nu}=-\frac{1}{2\sqrt{-g}}[\upsilon_{\sigma}\epsilon^{\sigma\mu\alpha\beta}\nabla_{\alpha}R^{\nu}_{\beta}+
\frac{1}{2}\upsilon_{\sigma\tau}\epsilon^{\sigma\nu\alpha\beta}R^{\tau\mu}_{\alpha\beta}]+(\mu\longleftrightarrow\nu).
\end{eqnarray}
Here,
$\upsilon_{\sigma}\equiv\nabla_{\sigma}\Theta$ and $\upsilon_{\sigma\tau}\equiv\nabla_{\sigma}\nabla_{\tau}\Theta$.
The energy-momentum tensor $T_{\mu\nu}$ is consists of the matter part $T^{m}_{\mu\nu}$ and the external field part $T^{\Theta}_{\mu\nu}$, defined respectively as
\begin{eqnarray}
T^{m}_{\mu\nu}&=&(\rho+p)U_\mu U_\nu-p g_{\mu\nu},\\
T^{\Theta}_{\mu\nu}&=&\beta(\partial_{\mu}\Theta)(\partial_{\nu}\Theta)-\frac{\beta}{2}g_{\mu\nu}(\partial^\lambda\Theta)(\partial_{\lambda}\Theta),
\end{eqnarray}
where $\rho$ is energy density, $p$ is pressure and $U$ is the four-vector velocity in co-moving
coordinates of the spacetime.
\section{HDE Model in CS Modified Gravity}
Granda and Oliveros \cite {[26]} proposed an infrared cut-off for the HDE which is the sum of the square
of the Hubble scale parameter and its time derivative given by
\begin{eqnarray}
\rho_{\Lambda}=3M^{2}_{P}(\alpha H^{2}+\beta \dot{H}),
\end{eqnarray}
where $\alpha$ and $\beta$ are constants which satisfy the restrictions of observational data and $H=\frac{\dot{a}}{a}$ is Hubble parameter.
Now, we discuss the FRW universe defined by line element
\begin{eqnarray}
ds^{2}=-dt^{2}+a^{2}(t)[\frac{dr^{2}}{1-\kappa r^{2}}+r^{2}(d\theta^{2}+\sin^{2}{\theta}d\phi^{2})].
\end{eqnarray}
where $\kappa$ is the curvature of the space. Here $\kappa=-1,0,1$ denotes open, flat and closed universe respectively.
The 00-component of the field equation (2) of FRW metric  and using Eq.(8), turns out to be
\begin{eqnarray}
H^{2}+\frac{\kappa}{a^{2}}= (\alpha H^{2}+\beta \dot{H})+\frac{1}{6}\dot{\Theta}^{2}.
\end{eqnarray}
Now we calculate the value of $\Theta$ by using the Eq.(3).
As for FRW universe the Pontryagin term $ ^{*}RR= {{^{*}R^a}_b}^{cd} {R^b}_{acd}$  vanishes, so
Eq.(3) takes the form
\begin{eqnarray}
g^{\mu \nu}\nabla_{\mu} \nabla_{\nu}\Theta= g^{\mu \nu}[\partial_{\mu}\partial_{\nu}\Theta - \Gamma^{\tau}_{\mu \nu}\partial_{\tau}\Theta]=0.
\end{eqnarray}
The solution of this equation can be found, in terms of $\dot{\Theta}$, as
\begin{eqnarray}
\dot{\Theta}=C a^{-3}.
\end{eqnarray}
Substituting the value of $\dot{\Theta}$ in Eq.(9) along with assumption $x= \ln a$, we arrive at
\begin{eqnarray}
\frac{dH^{2}}{dx}+\frac{2(\alpha-1)}{\beta}H^{2}+\frac{C^{2}}{3\beta}e^{-6x}-\frac{2\kappa}{\beta}e^{-2x}=0.
\end{eqnarray}
The solution  of this differential equation is obtained using direct integration technique, given as
\begin{eqnarray}
H^{2}(x)= C_{1}e^{\frac{-2(\alpha-1)}{\beta}x}-\frac{C^{2}}{6(\alpha-3\beta-1)}e^{-6x}+\frac{\kappa}{\alpha-\beta-1}e^{-2x}.
\end{eqnarray}
The conservation equation is given by \cite{[26]}
\begin{eqnarray}
\dot{\rho}_{\Lambda}+3H(\rho_{\Lambda}+p_{\Lambda})=0.
\end{eqnarray}
The holographic energy density and pressure density are related by the barotropic  equation of state (EoS) defined as
$p_{\Lambda}=\omega_{\Lambda}\rho_{\Lambda}$, where $\omega_{\Lambda}$ is the EoS parameter. Then the last equation takes the form
\begin{eqnarray}
\omega_{\Lambda}=-1-\frac{2\alpha \dot{H}+\beta \frac{\ddot{H}}{H}}{3(\alpha H^{2}+\beta \dot{H})}.
\end{eqnarray}
Using Eq.(13) in Eq.(15), we have
\begin{eqnarray}
\omega_{\Lambda}=-\frac{1}{3}(\frac{\frac{(3\beta-2\alpha+2)}{\beta}C_{1}e^{-\frac{2(\alpha-1)x}{\beta}}+\frac{\kappa(\alpha-\beta)}{(\alpha-\beta-1)}e^{-2x}+
\frac{C^{2}(\alpha-3\beta)}{2(\alpha-3\beta-1)}e^{-6x}} {C_{1}e^{-\frac{2(\alpha-1)x}{\beta}}+\frac{\kappa(\alpha-\beta)}{(\alpha-\beta-1)}e^{-2x}-
\frac{C^{2}(\alpha-3\beta)}{6(\alpha-3\beta-1)}e^{-6x}})
\end{eqnarray}
It is mentioned here that the EoS parameter is time dependent that can be transit from $\omega_{\Lambda}>-1$ to $\omega_{\Lambda}<-1$ \cite{[24]}.
Although, the recent studies \cite{[25]} of DE properties are mildly support the models with $\omega_{\Lambda}$ crossing $-1$.
For the flat case, when $\kappa=0$, by using the assumption $\alpha=3\beta$, the last equation turns out to be
\begin{eqnarray}
\omega_{\Lambda}=\frac{\alpha-2}{\alpha},
\end{eqnarray}
which describes the EoS parameter in term of constant $\alpha$ only.
The accelerated expansion of the universe can be obtained with restrictions on $\alpha$ such that
$1<\alpha<\frac{3}{2}$, if the phase $ -1<\omega_{\Lambda}<-\frac{1}{3}$ is under consideration.
If we consider $0<\alpha<1$ then the holographic energy and pressure density illustrates phantom-like theory of the
evolution alongwith $\omega_{\Lambda}<-1$.
\section{Correspondence Between Holographic and Scalar Field Models}
Here we establish a correspondence between infrared cut-off proposed by Granda and Oliveros \cite{[26]} for the holographic dark energy
density and some of famous scalar field models, like quintessence model, tachyon model, K-essence model and dilaton model.
We compare the holographic density defined by Gronda and Oliveros with the density of corresponding scalar field model.
Further, we equate the barotropic EoS parameter, given in Eq.(17), with the EoS parameter of the
corresponding scalar field models to find the scalar field and the potential energy.
\subsection{Quintessence Model in CS Modified Gravity}
Quintessence is  described as canonical scalar field which was purposed to investigate the late-time cosmic acceleration.
The pressure density and energy density of quintessence scalar field are defined as
\begin{eqnarray}
p_{\phi}=\frac{1}{2}\dot{\phi}^{2}-V(\phi),\\
\rho_{\phi}=\frac{1}{2}\dot{\phi}^{2}+V(\phi),
\end{eqnarray}
where the dot denotes derivative with respect to $t$.
The dark energy EoS parameter for the quintessence scalar field is
\begin{eqnarray}
\omega_{\phi}=\frac{\frac{1}{2}\dot{\phi}^{2}-V(\phi)}{\frac{1}{2}\dot{\phi}^{2}+V(\phi)}.
\end{eqnarray}
Now, we compare the new HDE modal $\omega_{\Lambda}$, given in Eq.(16), with that of quintessence DE modal
$\omega_{\phi}$, given in Eq.(20), and obtain
\begin{eqnarray}
-\frac{1}{3}(\frac{\frac{(3\beta-2\alpha+2)}{\beta}C_{1}e^{-\frac{2(\alpha-1)x}{\beta}}+\frac{\kappa(\alpha-\beta)}{(\alpha-\beta-1)}e^{-2x}+
\frac{C^{2}(\alpha-3\beta)}{2(\alpha-3\beta-1)}e^{-6x}} {C_{1}e^{-\frac{2(\alpha-1)x}{\beta}}+\frac{\kappa(\alpha-\beta)}{(\alpha-\beta-1)}e^{-2x}-
\frac{C^{2}(\alpha-3\beta)}{6(\alpha-3\beta-1)}e^{-6x}})=\frac{\frac{1}{2}\dot{\phi}^{2}-V(\phi)}{\frac{1}{2}\dot{\phi}^{2}+V(\phi)}
\end{eqnarray}
On comparing Eq.(7) and Eq.(19), it comes out that
\begin{eqnarray}
\frac{1}{2}\dot{\phi}^{2}+V(\phi)=3M^{2}_{P}(\alpha H^{2}+\beta \dot{H}).
\end{eqnarray}
Making use of Eq.(22) in Eq.(21) yields the explicit expression for  $\dot{\phi}$ and potential $V(\phi)$
\begin{eqnarray}
\dot{\phi}^{2}=2M_{p}^{2}[\frac{(\alpha-1)}{\beta}C_{1}e^{-\frac{2(\alpha-1)x}{\beta}}+\frac{\kappa(\alpha-\beta)}{(\alpha-\beta-1)}e^{-2x}+
\frac{C^{2}(\alpha-3\beta)}{2(\alpha-3\beta-1)}e^{-6x}]
\end{eqnarray}
and
\begin{eqnarray}
V(\phi)= M_{P}^{2}[\frac{C_{1}(3\beta-\alpha+1)}{\beta}e^{-\frac{2(\alpha-1)}{\beta}x}+\frac{\kappa(\alpha-\beta)}{\alpha-\beta-1}e^{-2x}]
\end{eqnarray}
respectively. As we assumed $x=\ln a$, it follows that $\dot{\phi}=\phi^{\prime}H$, where prime denotes the derivative with respect to $x$.
On substituting this value, Eq.(23) turns out to be
\begin{eqnarray}
\phi^{\prime}=\sqrt{2}M_{P}[\frac{\frac{(\alpha-1)}{\beta}C_{1}e^{-\frac{2(\alpha-1)x}{\beta}}+\frac{\kappa(\alpha-\beta)}{(\alpha-\beta-1)}e^{-2x}+
\frac{C^{2}(\alpha-3\beta)}{2(\alpha-3\beta-1)}e^{-6x}}{ C_{1}e^{\frac{-2(\alpha-1)}{\beta}x}+\frac{\kappa}{\alpha-\beta-1}e^{-2x}-\frac{C^{2}}{6(\alpha-3\beta-1)}e^{-6x}}]^{\frac{1}{2}}.
\end{eqnarray}
For the flat case, i.e., $\kappa=0$, using the assumption $\alpha=3\beta$  and taking $\phi(t_{0})=0$ at initial time $t_{0}=0$, the Eqs.(25) and (24) become
\begin{eqnarray}
\phi(t)=\sqrt{\frac{6(\alpha-1)}{\alpha}}M_{P} \ln{t}
\end{eqnarray}
and
\begin{eqnarray}
V(\phi)=\frac{3}{\alpha}C_{1}M_{P}^{2} e^{-\sqrt{\frac{6(\alpha-1)}{\alpha}}\frac{\phi}{M_{P}}}.
\end{eqnarray}
The potential $V(\phi)$ becomes a source of accelerated expansion of the universe if $\alpha<\frac{3}{2}$.
If we discuss the phase-space analysis, the potential $V(\phi)$ corresponding to scalar field $\phi$
behaves like an attractor solution which is indication of accelerated expansion for $\alpha<\frac{3}{2}$,
same conditions are followed in power law accelerated expansion.
The detailed analysis of dynamics of an exponential potential $V(\phi)$ is given in \cite{[27]}.

\subsection{New Holographic Tachyon Model in CS Modified Gravity}
The tachyon modal is considered as a good candidate for dark energy.
The idea of tachyon is $40$ years old and attained much attention again after the research papers by Sen \cite{[22]}.
The tachyon scalar field $\phi$ is studied with Born-Infeld Lagrangian
$V(\phi)\sqrt{1-g^{\mu\nu}\partial_{\mu}\phi \partial_{\nu}\phi}$ which have minimal coupling with gravity.
In the tachyon model the energy and pressure densities are given by
\begin{eqnarray}
\rho_{T}=\frac{V(\phi)}{\sqrt{1-\dot{\phi}^{2}}},\\
P_{T}=V(\phi)\sqrt{1-\dot{\phi}^{2}}.
\end{eqnarray}
The EoS parameter for tachyon scalar field is
\begin{eqnarray}
\omega_{T}=\dot{\phi}^{2}-1.
\end{eqnarray}
The comparison between baroscopic EoS given in Eq.(16) and  Eq.(30), yields
\begin{eqnarray}
1-\dot{\phi}^{2}=\frac{1}{3}(\frac{\frac{(3\beta-2\alpha+2)}{\beta}C_{1}e^{-\frac{2(\alpha-1)x}{\beta}}+\frac{\kappa(\alpha-\beta)}{(\alpha-\beta-1)}e^{-2x}+
\frac{C^{2}(\alpha-3\beta)}{2(\alpha-3\beta-1)}e^{-6x}} {C_{1}e^{-\frac{2(\alpha-1)x}{\beta}}+\frac{\kappa(\alpha-\beta)}{(\alpha-\beta-1)}e^{-2x}-
\frac{C^{2}(\alpha-3\beta)}{6(\alpha-3\beta-1)}e^{-6x}}),
\end{eqnarray}
which implies that
\begin{eqnarray}
\dot{\phi}^{2}=\frac{2}{3}[\frac{C_{1}\frac{\alpha-1}{\beta}e^{-\frac{2(\alpha-1)x}{\beta}}+\frac{\kappa(\alpha-\beta)}{(\alpha-\beta-1)}e^{-2x}-
\frac{C^{2}(\alpha-3\beta)}{2(\alpha-3\beta-1)}e^{-6x}}{C_{1}e^{-\frac{2(\alpha-1)x}{\beta}}+\frac{\kappa(\alpha-\beta)}{(\alpha-\beta-1)}e^{-2x}-
\frac{C^{2}(\alpha-3\beta)}{6(\alpha-3\beta-1)}e^{-6x}}].
\end{eqnarray}
Since $\dot{\phi}=\phi^{\prime}H$ and using corresponding values of $\phi$ and $H$, the evolutionary form of tachyon scalar field is yield as
\begin{eqnarray}
&&\phi(a)-\phi(0)\nonumber\\
&=&\int_{0}^{\ln a}{\frac{1}{H}\sqrt{\frac{2(C_{1}\frac{\alpha-1}{\beta}e^{-\frac{2(\alpha-1)x}{\beta}}+\frac{\kappa(\alpha-\beta)}{(\alpha-\beta-1)}e^{-2x}-
\frac{C^{2}(\alpha-3\beta)}{2(\alpha-3\beta-1)}e^{-6x})}{3(C_{1}e^{-\frac{2(\alpha-1)x}{\beta}}+\frac{\kappa(\alpha-\beta)}{(\alpha-\beta-1)}e^{-2x}-
\frac{C^{2}(\alpha-3\beta)}{6(\alpha-3\beta-1)}e^{-6x})}}}dx.
\end{eqnarray}
The analytic solution of this integral cannot be found explicitly.
For approximate solution, assume that $\alpha=3\beta$ and $\phi(0)=0$, i.e., at initial time and we consider the flat universe, i,e. $\kappa=0$, the last equation becomes
\begin{eqnarray}
\phi(a)=\sqrt{2(1-\frac{1}{\alpha})}\int_{0}^{\ln a}\frac{dx}{\sqrt{C_{1}e^{\frac{-6(\alpha-1)x}{\alpha}}+\frac{C^{2}}{6}e^{-6x}}}.
\end{eqnarray}
The integral on R.H.S can be solved in term of hypergeometric function as
\begin{eqnarray}
\phi(a)=2\sqrt{\frac{\alpha-1}{3\alpha}}\frac{e^{3x}}{C}~ _{2}F_{1}[\frac{1}{2}, \frac{\alpha}{2}, 1+\frac{\alpha}{2}, -\frac{6e^{\frac{6x}{\alpha}}C_{1}}{C^{2}}].
\end{eqnarray}
Finally, by substituting $x=\ln{a}$, we obtain
\begin{eqnarray}
\phi(a)=2\sqrt{\frac{\alpha-1}{3\alpha}}\frac{a^{3}}{C}~ _{2}F_{1}[\frac{1}{2}, \frac{\alpha}{2}, 1+\frac{\alpha}{2}, -\frac{6a^{\frac{6}{\alpha}}C_{1}}{C^{2}}].
\end{eqnarray}
Clearly, for $\alpha=1$, it yields
\begin{eqnarray}
\phi(a)=0.
\end{eqnarray}
Now, the comparison between Gronda and Oliveros cut-off, given in Eq.(7), and holographic tachyon model density, given in Eq.(28), yields
\begin{eqnarray}
\rho_{\Lambda}=3M_{P}^{2}(\alpha H^{2}+\beta \dot{H})=\frac{V(\phi)}{\sqrt{1-\dot{\phi}^{2}}}.
\end{eqnarray}
Substituting the values of $\phi$ and $H$, the tachyon potential energy turns to be
\begin{eqnarray}
V(\phi)=\sqrt{3}M_{P}^{2}\sqrt{[C_{1}e^{-\frac{2(\alpha-1)x}{\beta}}+\frac{\kappa(\alpha-\beta)}{(\alpha-\beta-1)}e^{-2x}-
\frac{C^{2}(\alpha-3\beta)}{6(\alpha-3\beta-1)}e^{-6x}]}\nonumber\\
\times \sqrt{[C_{1}\frac{3\beta-2\alpha+1}{\beta}e^{-\frac{2(\alpha-1)x}{\beta}}+\frac{\kappa(\alpha-\beta)}{(\alpha-\beta-1)}e^{-2x}-
\frac{C^{2}(\alpha-3\beta)}{2(\alpha-3\beta-1)}e^{-6x}]}.
\end{eqnarray}
Again, for  $\kappa=0$ and $\alpha=3\beta$, we have
\begin{eqnarray}
V(\phi)=3M_{P}^{2}C_{1}\sqrt{\frac{2-\alpha}{\alpha}}e^{6(\frac{1-\alpha}{\alpha})x}
\end{eqnarray}
For particular case $\alpha=1$, the potential $V(\phi)$ is constant which corresponds to ghost condensate scenario discussed in \cite{[31]}.

\subsection{New Holographic K-essence Modal in CS Modified Gravity}
The concept of k-essence scalar field model was introduced by Armendariz and Mukhanov \cite{[23]} to
explain the accelerated expansion of the universe. The theory of k-essence deals with dynamical attractor
solutions which act as a cosmological constant. The scalar field action for K-essence modal is defined as
\begin{eqnarray}
S=\int d^{4}x\sqrt{-g}p(\phi,X),
\end{eqnarray}
where $p(\phi, X)$ denotes pressure density and most of time it corresponds to Lagrangian density defined as
$p(\phi, X)=f(\phi)\psi(X)$.
In string theory, the Lagrangian density is transformed into
\begin{eqnarray}
p(\phi,X)=f(\phi)(-X+X^{2}).
\end{eqnarray}
The energy density of the field $\phi$ corresponding to the Lagrangian density expression is given by
\begin{eqnarray}
\rho(\phi,X)=f(\phi)(-X+3X^{2}).
\end{eqnarray}
Using Eq.(42) and Eq.(43), one can easily obtain EoS parameter, given as
\begin{eqnarray}
\omega_{K}=\frac{X-1}{3X-1}.
\end{eqnarray}
In particular $X<\frac{3}{2}$ , the EoS $\omega_{\phi}<-\frac{1}{3}$ indicate the  accelerated expansion.
The comparison  between  Eq.(16) and new EoS parameter Eq.(44) yields
\begin{eqnarray}
X=\frac{1}{3}[\frac{\frac{3\beta-\alpha+1}{\beta}C_{1}e^{-\frac{2(\alpha-1)x}{\beta}}+\frac{2\kappa(\alpha-\beta)}{(\alpha-\beta-1)}e^{-2x}}
{\frac{2\beta-\alpha+1}{\beta}C_{1}e^{-\frac{2(\alpha-1)x}{\beta}}+\frac{\kappa(\alpha-\beta)}{(\alpha-\beta-1)}e^{-2x}
+\frac{C^{2}(\alpha-3\beta)}{6(\alpha-3\beta-1)}e^{-6x}}]
\end{eqnarray}
The term  $\dot{\phi}^{2}=2X$ defined in \cite{[12]} and $\dot{\phi}=\phi^{\prime}H$, using these expressions, the evolutionary form of K-essence scalar field takes the form
\begin{eqnarray}
&&\phi(a)-\phi(0)=\sqrt{\frac{2}{3}}\times\nonumber\\
&&\int_{0}^{\ln a}{\frac{1}{H}\sqrt{\frac{\frac{3\beta-\alpha+1}{\beta}C_{1}e^{-\frac{2(\alpha-1)x}{\beta}}+\frac{2\kappa(\alpha-\beta)}{(\alpha-\beta-1)}e^{-2x}}
{\frac{2\beta-\alpha+1}{\beta}C_{1}e^{-\frac{2(\alpha-1)x}{\beta}}+\frac{\kappa(\alpha-\beta)}{(\alpha-\beta-1)}e^{-2x}
+\frac{C^{2}(\alpha-3\beta)}{6(\alpha-3\beta-1)}e^{-6x}}}}dx.
\end{eqnarray}
The compassion of Eq.(16) and Eq.(44), alongwith values of $H$, yields the expression for $f(\phi)$ as
\begin{eqnarray}
f(\phi)&=&\frac{3M_{P}^{2}(1-3\omega_{\Lambda})^{2}}{2(1-\omega_{\Lambda})}\nonumber\\
&\times &[C_{1}e^{-\frac{2(\alpha-1)x}{\beta}}+\frac{\kappa(\alpha-\beta)}{(\alpha-\beta-1)}e^{-2x}
-\frac{C^{2}(\alpha-3\beta)}{6(\alpha-3\beta-1)}e^{-6x}],
\end{eqnarray}
which can be further written as
\begin{eqnarray}
f(\phi)=\frac{M_{P}^{2}}{2}\frac{[\frac{2(2\beta-\alpha+1)}{\beta}C_{1}e^{-\frac{2(\alpha-1)x}{\beta}}+\frac{2\kappa(\alpha-\beta)}{(\alpha-\beta-1)}e^{-2x}
+\frac{C^{2}(\alpha-3\beta)}{3(\alpha-3\beta-1)}e^{-6x}]^{2}}{\frac{2\alpha-2}{\beta}C_{1}e^{-\frac{2(\alpha-1)x}{\beta}}+\frac{2\kappa(\alpha-\beta)}
{(\alpha-\beta-1)}e^{-2x}}.
\end{eqnarray}
Solving the Eqs.(45),(46) and (48) analytically, we consider the flat case, i,e. $\kappa=0$, use the assumption $\alpha=3\beta$ and $\phi(0)=0$ for the initial time $t_{0}=0$, it turns out to be
\begin{eqnarray}
X&=&\frac{1}{3-\alpha},\\
\phi(a)&=&\sqrt{\frac{2}{3-\alpha}}{\frac{e^{3x}}{C}~ _{2}F_{1}[\frac{1}{2},\frac{\alpha}{2},1+\frac{\alpha}{2},\frac{-6e^{\frac{6x}{\alpha}}}{C^{2}}C_1]},\\
f(\phi)&=&\frac{M_{P}^{2}}{3}\frac{(3-\alpha)^{2}}{\alpha(\alpha-1)}C_{1}e^{-\frac{6(\alpha-1)}{\alpha}}x.
\end{eqnarray}
For $\alpha=\frac{3}{2}$ the above equations turned into
\begin{eqnarray}
X&=&\frac{2}{3},\\
\phi(a)&=&\sqrt{\frac{4}{3}}{\frac{a^{3}}{C}~ _{2}F_{1}[\frac{1}{2},\frac{3}{4},\frac{7}{4},\frac{-6a^{6}}{C^{2}}C_1]},\\
f(\phi)&=&M_{P}^{2}C_{1}a^{-2}.
\end{eqnarray}
The potential $V(\phi)$ obey the power law expansion analysed in \cite{[27]}.

\subsection{New Holographic Dilaton Field in CS Modified\\ Gravity}
The dilaton model of DE is described by a 4-dimensional effective low-energy limit action in string theory.
It includes the higher order kinetic energy term which may be negative in the framework of Einstein relativity.
It indicates that the dilaton model works like a phantom-type scalar field.
The dilaton scalar field  model is defined by the pressure  density
\begin{eqnarray}
P_{d}=-X+c_{1}e^{\lambda \phi}X^{2},
\end{eqnarray}
where $c_{1}$ and $\lambda$ are positive constants.
The corresponding dilaton energy density is given by
 \begin{eqnarray}
\rho_{d}=-X+3c_{1}e^{\lambda \phi}X^{2},
\end{eqnarray}
where $ 2X=\dot{\phi}^{2}$. The EoS parameter $\omega_{d}=\frac{P_{d}}{\rho_{d}}$ can be obtained from Eqs. (55) and (56).
\begin{eqnarray}
\omega_{d}=\frac{-1+c_{1}e^{\lambda \phi}X}{-1+3c_{1}e^{\lambda \phi}X}.
\end{eqnarray}
Now, we compare Eq.(57) with new holographic EoS parameter, given in Eq.(16), i.e., $\omega_{d}=\omega_{\Lambda}$ to obtain
\begin{eqnarray}
C^{\prime}e^\frac{\lambda \phi}{2}X=\frac{1}{3}[\frac {\frac{2\alpha-2}{\beta}C_{1}e^{-\frac{2(\alpha-1)x}{\beta}}+\frac{2\kappa(\alpha-\beta)}
{(\alpha-\beta-1)}e^{-2x}}{\frac{2(2\beta-\alpha+1)}{\beta}C_{1}e^{-\frac{2(\alpha-1)x}{\beta}}+\frac{2\kappa(\alpha-\beta)}{(\alpha-\beta-1)}e^{-2x}
+\frac{C^{2}(\alpha-3\beta)}{3(\alpha-3\beta-1)}e^{-6x}}].
\end{eqnarray}
Making use of $X=\frac{\dot{\phi}^{2}}{2}$ and $\dot{\phi}=\phi^{\prime}H$ in the last equation and then integrating with respect to $x$, we get
\begin{eqnarray}
&& e^{\frac{\lambda \phi(a)}{2}}-e^{\lambda \phi(0)}=\frac{\lambda}{\sqrt{6C^{\prime}}}\times \nonumber\\
&&\int_{0}^{\ln a}{\frac{1}{H}\sqrt{\frac{\frac{(2-2\alpha)}{\beta}C_{1}e^{-\frac{2(\alpha-1)x}{\beta}}+\frac{2\kappa(\alpha-\beta)}{(\alpha-\beta-1)}e^{-2x}}
{\frac{2(2\beta-\alpha+1)}{\beta}C_{1}e^{-\frac{2(\alpha-1)x}{\beta}}+\frac{\kappa(\alpha-\beta)}{(\alpha-\beta-1)}e^{-2x}
+\frac{C^{2}(\alpha-3\beta)}{3(\alpha-3\beta-1)}e^{-6x}}}}dx
\end{eqnarray}
To obtain the evolutionary form of dilaton field, we consider the flat universe, i.e.,  $\kappa=0$, using assumption $\alpha=3\beta$ and at initial time $t_{0}=0$  the $\phi(0)=0$ and have
\begin{eqnarray}
\phi(a)=\frac{\lambda}{6C^{\prime}}\sqrt{\frac{1-\alpha}{3-\alpha}}e^{\frac{6x}{C^{2}}}
~_{2}F_{1}[\frac{1}{2},\frac{\alpha}{2},1+\frac{\alpha}{2},\frac{-6e^{\frac{6x}{\alpha}}}{C^{2}}C_1].
\end{eqnarray}
Re-substituting $x=\ln{a}$ in the last equation, we arrived at
\begin{eqnarray}
\phi(a)=\frac{\lambda}{\sqrt{6C^{\prime}}}\sqrt{\frac{1-\alpha}{3-\alpha}}a^{\frac{6}{C^{2}}}
~_{2}F_{1}[\frac{1}{2},\frac{\alpha}{2},1+\frac{\alpha}{2},\frac{-6a^{\frac{6}{\alpha}}}{C^{2}}C_1],
\end{eqnarray}
which is in term of hypergeometric function.
\section{Conclusion}
The accelerated expansion of the universe is a most discussed issue in the recent past.
In this paper, we found the EoS parameter $\omega_{\Lambda}$ which describe the accelerated
expansion of universe under certain restrictions on the parameter $\alpha$. It is shown that for the accelerated expansion phase
$ -1<\omega_{\Lambda}<-\frac{1}{3}$, the parameter  $\alpha$ varies  according as $1<\alpha<\frac{3}{2}$. Furthermore, for $0<\alpha<1$, the holographic energy
and pressure density illustrates phantom-like theory of the evolution when $\omega_{\Lambda}<-1$.

We explored the scalar field $\phi$ and potential $V(\phi)$ of different holographic dark energy models such
that quintessence, techyon, K-essence and dilaton.
The potential $V(\phi)$ becomes a source of accelerated expansion of the universe if $\alpha<\frac{3}{2}$.
When we discuss the phase-space analysis, the potential $V(\phi)$ corresponding to scalar field $\phi$
behaves like an attractor solution which is indication of accelerated expansion for $\alpha<\frac{3}{2}$,
same conditions are followed in power law accelerated expansion.
The detailed analysis of dynamics of an exponential potential $V(\phi)$ is given in \cite{[27]}.

\vspace{0.5cm}

{\bf Acknowledgment}
We acknowledge the remarkable assistance of the Higher Education
Commission Islamabad, Pakistan, and thankful for its financial
support through the {\it Indigenous PhD 5000 Fellowship Program
Batch-III}.

\end{document}